\documentclass[prl,aps,twocolumn,showpacs,nobibnotes]{revtex4}
\usepackage{graphicx}% Include figure files
\usepackage{dcolumn}% Align table columns on decimal point
\usepackage{bm}% bold math
\usepackage{SIunits}

\begin{document}

\title{Distinct electronic nematicities between electron and hole underdoped iron pnictides}
\author{J. J. Ying$^{1}$, X. F. Wang$^{1}$, T. Wu$^{1}$, Z. J. Xiang$^{1}$, R. H. Liu$^{1}$, Y. J. Yan$^{1}$, A. F. Wang$^{1}$, M. Zhang$^{1}$, G. J. Ye$^{1}$,
P. Cheng$^{1}$, J. P. Hu$^{2}$ and X. H. Chen$^{1}$}
\altaffiliation{Corresponding author} \email{chenxh@ustc.edu.cn}
\affiliation{$^{1}$Hefei National Laboratory for Physical Science at
Microscale and Department of Physics, University of Science and
Technology of China, Hefei, Anhui 230026, People's Republic of
China\\$^{2}$Department of Physics, Purdue University, West
Lafayette, Indiana 47907, USA}

\begin{abstract}
We systematically investigated the in-plane resistivity anisotropy
of electron-underdoped $EuFe_{2-x}Co_xAs_2$ and
$BaFe_{2-x}Co_xAs_2$, and hole-underdoped $Ba_{1-x}K_xFe_2As_2$.
Large in-plane resistivity anisotropy was found in the former
samples, while {\it tiny} in-plane resistivity anisotropy was
detected in the latter ones. When it is detected, the anisotropy
starts above the structural transition temperature and increases
smoothly through it. As the temperature is lowered further, the
anisotropy takes a dramatic enhancement through the magnetic
transition temperature. We found that the anisotropy is universally
tied to the presence of non-Fermi liquid T-linear behavior of
resistivity. Our results demonstrate that the nematic state is
caused by electronic degrees of freedom, and  the microscopic
orbital involvement in magnetically ordered state must be
fundamentally different between the hole and electron doped
materials.

\end{abstract}


\vskip 300 pt

\maketitle

The newly discovered iron-based high temperature superconductors
provide a new family of materials to explore the mechanism of
high-$T_c$ superconductivity besides high-$T_c$ cuprates
superconductors\cite{Kamihara, chenxh, ren, rotter}. The parent
compounds undergo a tetragonal to orthorhombic structure transition
and a collinear antiferromagnetic (CAFM) transition. The structure
transition temperature($T_S$) is higher than the CAFM transition
temperature($T_N$) in 1111 system, while for the 122 parent compound
$T_N$ is almost the same with $T_S$. Superconductivity arises when
both transitions are suppressed via electron or hole doping. The
origin of the structure transition and CAFM transition is still
unclear, it was proposed that antiferromagnetic fluctuation
\cite{Fang,xu} or orbital ordering\cite{Kr¨¹ger,Lee, Valenzuela} may
play an important role in driving the transitions.

Recent works showed a large in-plane resistivity anisotropy below
$T_S$ or $T_N$ in Co-doped Ba122 system, though the distortion of
the orthorhombic structure is less than 1 \%\cite{Chu, Tanatar} in
the CAFM state. The resistivity is smaller along the
antiferromagnetic $a$ direction than along the ferromagnetic $b$
direction which is opposite to our intuitive thinking. Moreover, the
resistivity anisotropy grows with doping and approaches its maximum
close to the point where superconductivity emerges. Other
experiments also reported the existence of anisotropy in  many
different physical properties  in the CAFM state of iron pnictides
including magnetic exchange coupling\cite{Zhao}, Fermi surface
topology\cite{Shimojima} and local density
distribution\cite{Chuang}.   Electronic anisotropies have also been
observed in many other materials, including underdoped cuprates,
quantum Hall systems and $Sr_3Ru_2O_7$\cite{Lilly, Borzi}. In
cuprates, electronic anisotropy can not be explained by small
structural orthorhombicity\cite{Ando, Hinkov} and it has been
proposed the large electron anisotropy is due to the emergence of
electron nematic phase. This idea has been
  widely investigated in the other two systems as well. In iron-pnictides,
  the origin of the the electronic anisotropy is still
elusive because of their complicated electronic structures and the
presence of many different degrees of freedom.

In this paper, we systematically investigated the in-plane
resistivity anisotropy of electron-underdoped $EuFe_{2-x}Co_xAs_2$
and $BaFe_{2-x}Co_xAs_2$, and hole-underdoped $Ba_{1-x}K_xFe_2As_2$
systems. Large in-plane resistivity behavior was found below $T_S$
or $T_N$ in $EuFe_{2-x}Co_xAs_2$ and $BaFe_{2-x}Co_xAs_2$, while
in-plane anisotropy is dramatically decreased and barely observable
in $Ba_{1-x}K_xFe_2As_2$. The resistivity shows T-linear behavior
above the temperature wherever the resistivity anisotropy starts to
emerge. In the $Ba_{1-x}K_xFe_2As_2$, no T-linear behavior in the
resistivity is observed in the normal state. For
$EuFe_{2-x}Co_xAs_2$, the resistivity behaves very differently at
$T_S$ and $T_N$.  The in-plane anisotropy starts to emerge even
above $T_S$ and increases smoothly through it. As the temperature is
lowered further, it takes a dramatic enhancement through the
magnetic phase transition. Therefore, our results provide  direct
evidence ruling out the possibility that the anisotropy is caused by
lattice distortion. The dramatic difference of the anisotropies in
the electron and hole doped materials suggests that magnetism must
be orbital-selective in a way that the magnetism in the hole
underdoped iron-pnictides may stem mainly from the $d_{xy}$ orbitals
while  the magnetism in the electron underdoped ones attributes
mainly to the $d_{yz} $ and $d_{xz}$ orbitals.

High quality single crystals with nominal composition
$EuFe_{2-x}Co_{x}As_2$ (x=0 ,0.067 ,0.1 ,0.225) and
$Ba_{1-x}K_xFe_2As_2$ (x=0.1 ,0.18) were grown by self-flux method
as described elsewhere\cite{wang}. Many shinning plate-like single
crystals can be obtained. $EuFe_2As_2$ has not only a CAFM or
structural transition around 190 K, but also an antiferromagnetic
transition of $Eu^{2+}$ ions around 20 K\cite{Wu}. As Co-doping
increases, the SDW transition and structure transition were
gradually separated and suppressed like the $BaFe_{2-x}Co_xAs_2$
system. Resistivity reentrance behavior due to the antiferromagnetic
ordering of $Eu^{2+}$ spins is also observed at low temperature. The
detailed in-plane resistivity of $EuFe_{2-x}Co_{x}As_2$ can be found
elsewhere\cite{He, Ying}. $Ba_{0.9}K_{0.1}Fe_2As_2$
 and $Ba_{0.82}K_{0.18}Fe_2As_2$ undergoes structural transition and SDW
transition at around 128K and 113 K, respectively. It is difficult
to directly measure the in-plane resistivity anisotropy because the
material naturally forms structure domains below $T_S$. In order to
investigate the intrinsic in-plane resistivity anisotropy we
developed a mechanical cantilever device that is able to detwin
crystals similar to the Ref.8. Crystals were cut parallel to the
orthorhombic a and b axes so that the orthorhombic a(b) direction is
perpendicular (parallel) to the applied pressure direction. $\rho_a$
(current parallel to a) and $\rho_b$ (current parallel to b) were
measured on the same sample using standard 4-point configuration. We
can get the same result with Ref.8 in Co-doped Ba122 system as shown
in Fig.4(d) and (e).

Fig.1(a) shows the  temperature dependence of in-plane resistivity
with the current flowing parallel to the orthorhombic $b$ direction
(black) and orthorhombic $a$ direction (red) of the detwined
$EuFe_2As_2$ sample. Compounds with a small amount of Co-doping show
large in-plane anisotropy as displayed in Fig.1(b) and (c) for
x=0.067 and x=0.1, respectively. More specifically, we can find a
much more obvious upturn of $\rho_b$ around $T_N$ or $T_S$ compared
to the twined in-plane resistivity as the blue line shows while
$\rho_a$ shows no upturn and drops very rapidly with the temperature
decreases below $T_N$ or $T_S$. The behavior of $\rho_a$ is very
similar to the $EuFe_2As_2$ polycrystalline samples\cite{Zhi Ren}.
When the Co-doping increases, the upturn of $\rho_b$ at $T_N$ or
$T_S$ becomes much more sharper while the drop of $\rho_a$ at $T_N$
or $T_S$ gradually disappears and  turns into a small upturn at the
temperature slightly lower than $\rho_b$. For the sample of x=0.225,
no in-plane anisotropy was found due to the complete suppression  of
the  structural or magnetic transition as shown in Fig.1(d). The
different behavior of $\rho_a$ and $\rho_b$ observed here is very
similar  to the ones in other parent or electron underdoped
iron-pnictides\cite{Tanatar, Chu}.
\begin{figure}[t]
\centering
\includegraphics[width=0.5 \textwidth]{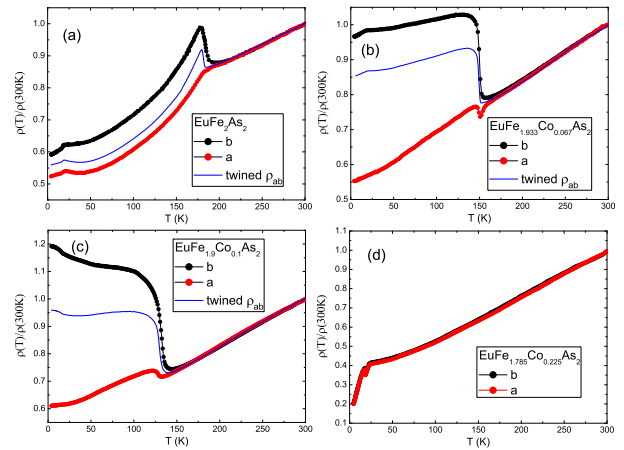}
\caption{(Color online) Temperature dependence of in-plane
resistivity with the electric current flow along $a$ direction (red)
and $b$ direction (black) respectively for the parent compound (a):
$EuFe_2As_2$, (b): $EuFe_{1.933}Co_{0.067}As_2$, (c):
$EuFe_{1.9}Co_{0.1}As_2$, (d): $EuFe_{1.775}Co_{0.225}As_2$. The
twined in-plane resistivity was also shown for comparison(blue
line).} \label{fig1}
\end{figure}

Further examining our data, we also notice that the structural
transition temperature $T_S$ and the magnetic transition temperature
$T_N$ become well separated as the Co-doping increases in
$EuFe_{2-x}Co_{x}As_2$. This separation provides us an opportunity
to investigate the effects of the structural transition and magnetic
transition on the $\rho_a$ and $\rho_b$ separately. The two
transition temperatures can be obtained by analyzing the heat
capacity of the samples. Treating the heat capacity in the x=0.225
sample as phonon background and subtracting it,  we obtain the
electronic part of the heat capacity of the x=0.067 and x=0.1
samples, $\Delta C_P$. Taking the x=0.067 sample as an example, the
$\Delta C_P$ has two distinct features as a function of temperature:
a very sharp peak at 150 K  and  a broad hump  around 157 K. Similar
to other isostructure materials, such as $BaFe_{2-x}Co_xAs_2$
\cite{Chu2}, we can attribute the sharp peak to the magnetic
transition and the broad hump to the structure transition. In the
x=0.067 sample, as shown in fig.2(a), $\rho_a$ shows a dip like
behavior from  $T_S$ to $T_N$. $\rho_a$ starts to drop rapidly at
the temperature coincident with $T_S$, and after reaching its
minimum value, it shows a very weak upturn at the temperature
coincident with $T_N$. $\rho_b$, however, has no observable feature
at $T_S$ but shows a large upturn around $T_N$. When the sample is
not detwinned, the in-plane resistivity $\rho_{ab}$ also shows a
large upturn.
 %around $T_N$. It was the first
%time we observed the two distinct behavior in one sample which
%indicated that the sudden decrease of $\rho_a$ is attributed to the
%structure transition, while the upturn of the resistivity is mainly
%attributed to the SDW transition. It is still unclear why the
%resistivity of some iron pnictide compounds shows upturn at $T_N$ or
%$T_S$ like $CaFe_2As_2$ while the resistivity of the others directly
%drop at $T_N$ or $T_S$ and no upturn was observed like
%$BaFe_2As_2$\cite{wu}. The observation of these two different
%resistivity response at $T_N$ and $T_S$ separately may explain the
%different resistivity response at $T_N$ or $T_S$ in iron pnictide.
For the x=0.1 sample, $T_S$ is suppressed to 140 K and $T_N$ is
suppressed to 130 K from the heat capacity measurement as shown in
Fig.2(b). $\rho_b$ starts to go upward above $T_S$ but $\rho_a$ does
not show any obvious response. The $d\rho_{ab}/d T$ curve of the
twined $\rho_{ab}$ peaks around $T_N$ while the peak of d$\rho_b/dT$
is  higher than $T_N$ and the one of $d\rho_a/dT$ is slightly lower
than $T_N$. With increasing the Co doping, the effect of the
structure transition on the $\rho_a$ is gradually wiped out which is
similar to $BaFe_{2-x}Co_xAs_2$ system, while the upturn behavior
$T_N$ becomes more noticeable.
%The effect on $\rho_b$ start to go upturn at the temperature
%slightly higher than $\rho_a$, which also indicate the anisotropic
%nature of the in-plane electron state.
\begin{figure}[t]
\centering
\includegraphics[width=0.5 \textwidth]{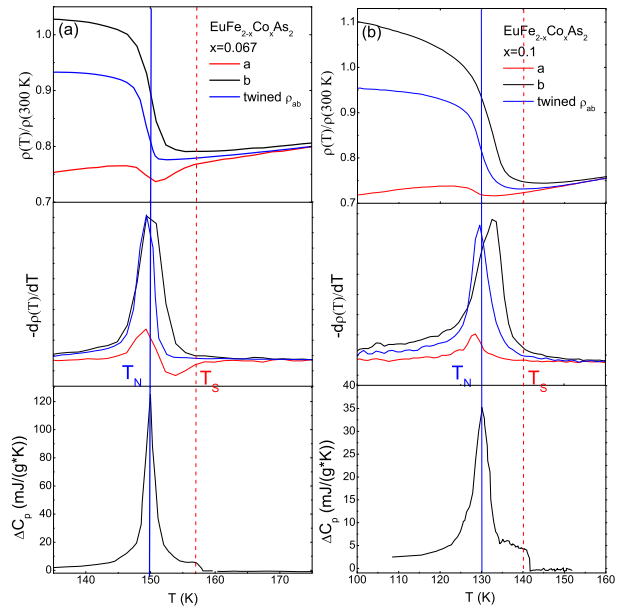}
\caption{(Color online) Temperature dependence of $\rho_a$, $\rho_b$
and twined $\rho_{ab}$, their differential curve and heat capacity
around the $T_S$ and $T_N$ for the sample of (a): x=0.067 and¡¡(b):
x=0.1. The red dashed line and blue solid line indicated $T_S$ and
$T_N$ respectively.} \label{fig2}
\end{figure}

We characterized the degree of in-plane resistivity anisotropy by
the ratio $\rho_b$/$\rho_a$. Fig.3(a) and (b) show in-plane
resistivity anisotropy $\rho_b$/$\rho_a$ and its related
differential curve for x=0.067 and x=0.1 samples, respectively. The
amplitude of the anisotropy is greatly increased with Co doping
though the magnetic transition and structure transition are
suppressed, similar to $BaFe_{2-x}Co_xAs_2$. The
in-plane resistivity anisotropy increases very rapidly at $T_N$ and
still gradually increases down to 4 K. A very sharp peak can be
observed in the differential curve of $\rho_b$/$\rho_a$ at the
temperature coincident with $T_N$. However, $\rho_b$/$\rho_a$ did
not show any obvious anomaly at $T_S$. The sharp increase of
in-plane resistivity anisotropy at $T_N$ indicates that in-plane
anisotropy is correlated to the magnetic transition rather than
structure transition. It also indicates that the in-plane electron
anisotropy is driven by the magnetic fluctuation or other hidden
electronic order rather than the small orthorhombic distortion.

\begin{figure}[t]
\centering
\includegraphics[width=0.5 \textwidth]{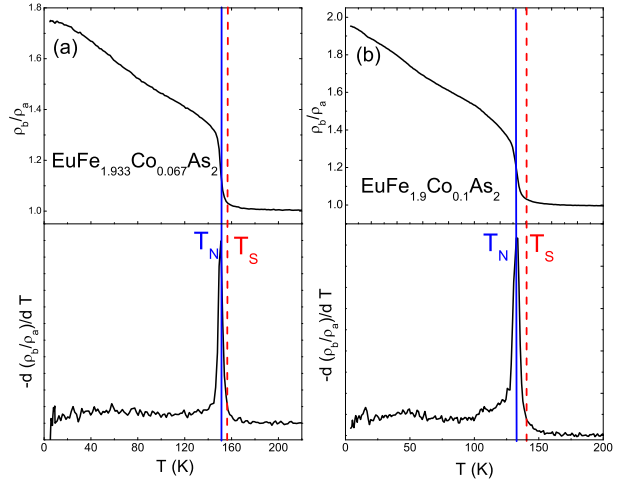}
\caption{(Color online) In-plane resistivity anisotropy
$\rho_b$/$\rho_a$ and its related differential curve for the sample
of (a): x=0.067 and (b): x=0.1. The blue solid line indicated $T_N$
and red dashed line indicated $T_S$.} \label{fig3}
\end{figure}
To understand the common feature of the in-plane resistivity
anisotropies in $EuFe_{2-x}Co_{x}As_2$ and $BaFe_{2-x}Co_xAs_2$, we
plot $\rho_a$ and $\rho_b$ in an enlarged temperature region as
shown in Fig.4(a), (b), (c), (d) and (e). For
$EuFe_{2-x}Co_{x}As_2$, it is very clear  that the in-plane
resistivity anisotropy emerges at temperatures higher than the
structure transition temperature as the black arrows indicate. It
suggests that fluctuations associated with the resistivity
anisotropy must emerge well above the $T_S$. Moreover,  T-linear
resistivity behaviors appear in a large temperature region above the
temperature  which $\rho_a$ and $\rho_b$ begin to show discrepancy.
We also observed this kind of feature in Co-doped Ba122 system as
shown in Fig.4(d), (e) for $BaFe_2As_2$ and
$BaFe_{1.83}Co_{0.17}As_2$ respectively. This non-Fermi liquid
behavior, or the T-linear behavior has also been observed in other
iron-pnictides, for example, $BaFe_2As_{2-x}P_x$ single crystals and
$SmO_{1-x}F_xFeAs$ polycrystaline samples\cite{Kasahara, Liu}. It
suggests that this behavior might be universal in electron underdoped
iron pnictides at high temperature. The T-linear resistivity
behavior  is also similar to the one in cuprates. However, in
$Ba_{2-x}K_{x}Fe_2As_2$, such a T-linear behavior is rapidly
suppressed through K doping. As shown in Fig.4(f) and (g),
$Ba_{0.9}K_{0.1}Fe_2As_2$ shows very tiny in-plane anisotropy
compared with its parent compound, $Ba_{0.82}K_{0.18}Fe_2As_2$
almost show no in-plane resistivity anisotropy and meanwhile the
normal state resistivity does not follow the T-linear behavior.  The
correspondence between the anisotropy and T-linear behavior of
resistivity presents in all of measured materials.  We did not find
a single exception.

Our above results have important implications for the origin of the
nematicity in iron-pnictides. First, our results strongly support
the nematic state in iron-pnictides is indeed an electronic nematic
state. Our measurements show that the in-plane resistivity
anisotropy  is closely related to the magnetic transition rather
than the structural transition and persists at temperature higher
than structural transition, suggesting the existence of nematicity
even in tetragonal lattice structure. Second, the distinct
anisotropy of resistivity between the hole and electron underdoped
materials reveals  the hidden interplay between magnetic and orbital
degrees of freedom. Recently, many theories focus on orbital
ordering which generates an unequal occupation of $d_{xz}$ and
$d_{yz}$ orbitals\cite{Chen} that breaks the rotational symmetry and
causes the lattice distortion\cite{Weicheng}. ARPES measurements
have provided orbital ordering evidence\cite{Shimojima}. Since there
is little anisotropy in the hole doped samples in their magnetically
ordered states, our results suggest that the resistivity anisotropy
is most likely induced  by orbital ordering rather than magnetic
ordering.  The orbital ordering  is suppressed rapidly by
hole-doping while it is relatively robust to electron-doping.  The
strong enhancement of the anisotropy around $T_N$ in electron doped
systems indicates that the magnetic ordering and orbital ordering
are intimately connected to each other in these systems. However,
our result on hole doped samples  suggest the magnetic ordering is
not simply a result of orbital ordering. Considering the facts that
the dominating orbitals are $t_{2g}$  and  the orbital ordering is
most likely due to the $d_{xz}$ and $d_{yz}$, we can conclude that
although the magnetically ordered states in both electron and hole
doped materials have an identical ordering wavevector, the two
states must differ microscopically  from their orbital involvements,
namely,  the magnetism is orbital selective in a way that the
magnetic ordering in hole-doped systems is attributed mostly from
the $d_{xy}$ orbital while the $d_{yz}$ and $d_{xz}$ make important
contributions to the magnetism in the electron doped ones. This
implication can be tested explicitly by ARPES experiments in hole
doped materials. Finally,  the correspondence between the anisotropy
and T-linear behavior  suggests the importance of electron-electron
correlation in causing the orbital and magnetic ordering. The
non-Fermi liquid behavior of T-linear resistivity can be understood
by the presence of strong orbital or magnetic fluctuations in all
three $t_{2g}$ orbitals. Consequently,  the suppression of the
non-fermi liquid behavior by hole-doping indicates the suppression
of orbital and magnetic fluctuations in the $d_{xz}$ and $d_{yz}$
ones.

\begin{figure}[t]
\centering
\includegraphics[width=0.5\textwidth]{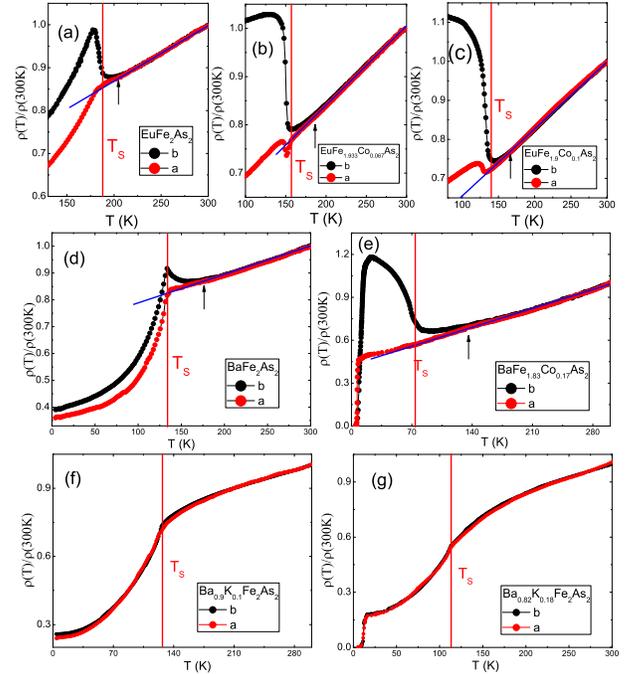}
\caption{(Color online)The temperature dependence of $\rho_a$ and
$\rho_b$ for (a)£º$EuFe_2As_2$, (b): $EuFe_{1.933}Co_{0.067}As_2$,
 (c): $EuFe_{1.9}Co_{0.1}As_2$,  (d): $BaFe_2As_2$,
(e): $BaFe_{1.83}Co_{0.17}As_2$, (f): $Ba_{0.9}K_{0.1}Fe_2As_2$  and
(g): $Ba_{0.82}K_{0.18}Fe_2As_2$. The black arrows indicate the
temperature where $\rho_a$ and $\rho_b$ begin to show discrepancy.
The red line indicates the $T_S$. The blue line is the linear fit of
the resistivity above the black arrow indicated temperature.}
\label{fig4}
\end{figure}

 %The
%in-plane resistivity of the parent and electron-doped iron pnictides
%at high temperature show T-linear behavior. The resistivity deviate
%from the T-linear behavior as soon as the in-plane electron
%anisotropy start to emerge at the temperature even higher than
%$T_S$. The T-linear resistivity behavior is also found in other iron
%pnictides and it is quite universal in the high Tc cuprates. These
%results indicate the similar normal state behavior between the two
%chasses of high $T_c$ superconductors. It is still unclear what is
%the origin of this unusual in-plane anisotropy. Some theory works
%based on orbital order had recently put forward to explain the
%unusual in-plane electron anisotropy. The orbital order can arise
%from the coupled spin-orbital degrees of freedom and lead to an
%unequal occupation of $d_{xz}$ and $d_{yz}$ orbitals\cite{Chen}. The
%emergent orbital order breaks the in-plane lattice symmetry and
%drives the orthorhombic lattice distortion\cite{Weicheng}. Recent
%ARPES works have shown that the parent 122 compounds show strongly
%orbital-dependent reconstruction of the electronic structure across
%the magnetostructural transition\cite{Shimojima} and the Fermi
%surfaces in AF state are dominantly composed of the $d_{xz}$
%orbital. The orbital characters of the Fermi surfaces may play an
%important role in this unusually in-plane resistivity anisotropy
%character.

In conclusion, we have measured the in-plane resistivity anisotropy
on electron-underdoped $EuFe_{2-x}Co_xAs_2$ and
$BaFe_{2-x}Co_xAs_2$, and hole-underdoped $Ba_{1-x}K_xFe_2As_2$
single crystals. Large in-plane resistivity anisotropy was found in
$EuFe_{2-x}Co_xAs_2$ which is quite similar to the isostructural
$BaFe_{2-x}Co_{x}As_2$ system, however anisotropy disappears in
hole-doped samples. We identified an universal correspondence
between the anisotropy and a T-linear behavior of resistivity at
high temperature. The different behavior of the anisotropy at $T_N$
and $T_S$ rules out the anisotropy is originated from the lattice
degree of freedom. The magnetic states in the hole and electron
doped systems are significantly different.

This work is supported by the Nature Science Foundation of China,
Ministry of Science and Technology and by Chinese Academy of
Sciences.


\begin{references}
\bibitem{Kamihara}
Y. Kamihara $et$ $al$., \emph{J. Am. Chem. Sco.} {\bf 130},
3296(2008).
\bibitem{chenxh}
X. H. Chen $et$ $al$., Nature {\bf 453}, 761(2008).
\bibitem{ren}
Z. A. Ren $et$ $al$., Europhys. Lett. {\bf 83}, 17002(2008).
\bibitem{rotter}
M. Rotter $et$ $al$., Phys. Rev. Lett. {\bf 101}, 107006(2008).
\bibitem{Kr¨¹ger}
F. Kr¨¹ger $et$ $al$., Phys. Rev. B {\bf 79}, 054504 (2009).
\bibitem{Fang}
C. Fang $et$ $al$., Phys. Rev. B {\bf 77}, 224509 (2008).
\bibitem{xu}
C. Xu $et$ $al$., Phys. Rev. B 78, 020501 (2008).
\bibitem{Chu}
Jiun-Haw Chu $et$ $al$., Science {\bf 329}, 824 (2010).
\bibitem{Tanatar}
M. A. Tanatar $et$ $al$., Phys. Rev. B {\bf 81}, 184508 (2010).
\bibitem{Ando}
Y. Ando $et$ $al$., Phys. Rev. Lett. {\bf 88}, 137005 (2002).
\bibitem{Hinkov}
V. Hinkov $et$ $al$., Science {\bf 319}, 597 (2008).
\bibitem{Lilly}
M. P. Lilly $et$ $al$., Phys. Rev. Lett. {\bf 82}, 394 (1999).
\bibitem{Borzi}
R. A. Borzi $et$ $al$., Science {\bf 315}, 214 (2007).
\bibitem{Zhao}
Jun Zhao $et$ $al$., Nature Physics {\bf 5}, 555 (2009).
\bibitem{Shimojima}
T. Shimojima $et$ $al$., Phys. Rev. Lett. {\bf 104}, 057002 (2010).
\bibitem{Chuang}
T.-M. Chuang $et$ $al$., Science {\bf 327}, 181 (2010).
\bibitem{Lee}
Chi-Cheng Lee $et$ $al$., Phys. Rev. Lett. {\bf 103}, 267001 (2009).
\bibitem{Valenzuela}
B. Valenzuela $et$ $al$., Phys. Rev. Lett. {\bf 105}, 207202(2010).
\bibitem{wang}
X. F. Wang  $et$ $al$.,  Phys. Rev. Lett. {\bf 102}, 117005(2009).
\bibitem{Wu}
T. Wu $et$ $al$., J. Magn. Magn. Mater. {\bf 321}, 3870 (2009).
\bibitem{He}
Y. He $et$ $al$., J. Phys.: Condens. Matter {\bf 22}, 235701 (2010).
\bibitem{Ying}
J. J. Ying $et$ $al$., Phys. Rev. B {\bf 81}, 052503 (2010).
\bibitem{Zhi Ren}
Zhi Ren $et$ $al$., Phys. Rev. B {\bf 78}, 052501 (2008).
\bibitem{wu}
G. Wu $et$ $al$., J. Phys.: Condens. Matter {\bf 20}, 422201 (2008).
\bibitem{Chu2}
Jiun-Haw Chu $et$ $al$., Phys. Rev. B {\bf 79}, 014506 (2009).
\bibitem{Kasahara}
S. Kasahara $et$ $al$., Phys. Rev. B {\bf 81}, 184519 (2010).
\bibitem{Liu}
R. H. Liu $et$ $al$., Phys. Rev. Lett {\bf 101}, 087001 (2008).
\bibitem{Chen}
C.-C. Chen $et$ $al$., Phys. Rev. B {\bf 82}, 100504(R) (2010).
\bibitem{Weicheng}
Weicheng Lv $et$ $al$., Phys. Rev. B {\bf 82}, 045125 (2010)



\newpage

\noindent

\end{references}
\end{document}